\documentclass[10pt]{iopart}
\usepackage[symbol]{footmisc}

\expandafter\let\csname equation*\endcsname\relax
\expandafter\let\csname endequation*\endcsname\relax
\usepackage{amsmath}
\usepackage{amsfonts}
\usepackage{xcolor}
\usepackage{graphicx}
\usepackage{hyperref}
\usepackage{cite}
\usepackage[normalem]{ulem}
\usepackage{bm}

\newcommand{\doiref}[2]{\href{https://doi.org/#1}{#2}}
\newcommand{\Pe}{\mathrm{Pe}}
\newcommand{\dd}{\mathop{}\!\mathrm{d}}

\DeclareMathOperator{\sech}{sech}

\begin{document}

\title[First-passage times of an active Ornstein--Uhlenbeck particle under resetting]{Persistence, resetting, and first-passage times of an active Ornstein--Uhlenbeck particle}

\author{Demosthenes K. Georgiou$^1$, Paul C. Bressloff$^{1,*}$ and Thibault Bertrand$^{1,*}$}

\address{$^{1}$Department of Mathematics, Imperial College London,
180 Queen's Gate, London SW7 2BZ, United Kingdom}

\address{$^*$Authors to whom any correspondence should be addressed.}

\ead{p.bressloff@imperial.ac.uk and t.bertrand@imperial.ac.uk}

%\keywords{Active Ornstein--Uhlenbeck particle, stochastic resetting, first-passage time, active matter}

\begin{abstract}
We study the first-passage time statistics of an active Ornstein--Uhlenbeck particle on a finite one-dimensional interval with an absorbing boundary at one end and a reflecting boundary at the other, subject to stochastic resetting. In the weak-activity regime, we develop a perturbative solution of the Fokker--Planck equation and use a renewal framework to obtain analytical expressions for the first-passage-time distribution and its low-order moments. In the absence of resetting, activity can either increase or decrease the mean first-passage time, depending on the relation between the persistence time and the diffusive time to absorption. When resetting is introduced, we determine both the onset of beneficial resetting and the finite range of resetting rates for which resetting lowers the mean first-passage time of the active system. We further show that the resetting transition depends on the velocity resetting protocol. Finally, by comparing the active particle with resetting to the passive Brownian particle without resetting, we construct a phase diagram that identifies the regions of parameter space in which activity and resetting together reduce the mean first-passage time.
\end{abstract}

%%%%%%%%%%%%%
% Section: Introduction  %
%%%%%%%%%%%%%
\section{Introduction}
First-passage processes describe the time taken for a stochastic system to reach a target for the first time. They find applications in many physical and biological settings, including diffusion-controlled reactions \cite{Benichou2014,Benichou2010}, animal search and foraging \cite{Bell1990,Fauchald2003,Benichou2005}, intracellular transport \cite{Bressloff2013}, and protein folding \cite{Polizzi2016,Lee2003}. Understanding the mechanisms that accelerate or delay first-passage events is therefore a central problem.

One such mechanism is stochastic resetting, which has emerged as a paradigm for modifying and controlling stochastic dynamics. Under resetting, the dynamics are randomly interrupted, and the system is returned to a predetermined state \cite{Evans2011a,Evans2020}. This repeated interruption and restart can give rise to behaviour that is absent in the underlying process, for example by breaking detailed balance and leading to the formation of non-equilibrium stationary states \cite{Eule2016,Evans2020}. In first-passage problems, resetting can make an otherwise divergent mean first-passage time (MFPT) finite and, in many cases, produce an optimal resetting rate that minimises the search time \cite{Evans2011a,Evans2011b,Reuveni2016,Pal2017}. While initial studies focused on the case of a single Brownian particle under Poissonian resetting, stochastic resetting has since been extended to a wide range of settings. These include more general resetting protocols \cite{Pal2016,Nagar2016}, interacting particle systems \cite{Nagar2023,Alston2025}, diffusion in switching environments \cite{Bressloff2020a,Bressloff2020b} and potential landscapes \cite{Pal2015}, and resetting with delays, such as finite return times \cite{MasoPuigdellosas2019a,Pal2019a,Bodrova2020,Pal2019b,Pal2020} or refractory periods \cite{Evans2019,MasoPuigdellosas2019b,Reuveni2014}. Resetting has also been studied in active-matter models \cite{Evans2020,Evans2018,Kumar2020}, where analytical results are typically harder to obtain because activity introduces additional dynamical variables beyond position.

Consequently, while the FPT statistics of Brownian particles with resetting have been studied extensively, much less is known about the combined effects of activity and resetting. Active particles provide minimal models of non-equilibrium systems whose constituents consume energy locally to self-propel \cite{Ramaswamy2010, Bechinger2016}, with examples ranging from bacteria and animal flocks to synthetic colloids and artificial swimmers \cite{Vicsek2012, Elgeti2015, Palacci2014}. Among the most widely studied active-particle models are run-and-tumble particles (RTPs), which are motivated by bacterial motion \cite{Berg2004} and alternate between persistent runs and random reorientation events, and active Brownian particles (ABPs), whose orientation decorrelates through rotational diffusion. Recent studies have focused on the FPT statistics with resetting for RTPs \cite{Santra2020,Tucci2022,Pal2024,Bressloff2025} and ABPs \cite{Scacchi2018,Baouche2025}.

Another important minimal model is the active Ornstein--Uhlenbeck particle (AOUP), in which the self-propulsion force has a Gaussian stationary distribution and an exponentially decaying time correlation \cite{Martin2021}. AOUPs have been used to study non-equilibrium steady states, collective phenomena such as motility-induced phase separation, and active particles in external potentials \cite{Szamel2014, Fodor2016}. Although their general dynamics under stochastic resetting have recently begun to be explored \cite{Shankari2025}, their FPT statistics in this context remain mostly unstudied to the best of our knowledge. For AOUPs, resetting not only returns the particle to a chosen position, but may also reinitialise the propulsion velocity. This simultaneous resetting of a kinematic variable (position) and a dynamic variable (velocity) is reminiscent of the physics of resetting in underdamped stochastic systems \cite{Gupta2019}. In confined geometries, this leads to a competition between passive diffusion, persistent self-propulsion, and restart. Consequently, the FPT depends not only on the resetting rate, but also on the particle's activity. Furthermore, because both active self-propulsion and stochastic resetting are inherently non-equilibrium processes, accelerating the search process inevitably carries a thermodynamic cost \cite{Fuchs2016,Busiello2020,Pal2023}, adding broader physical significance to identifying regimes of beneficial resetting.

\begin{figure}
    \centering
    \includegraphics[width=0.5\textwidth]{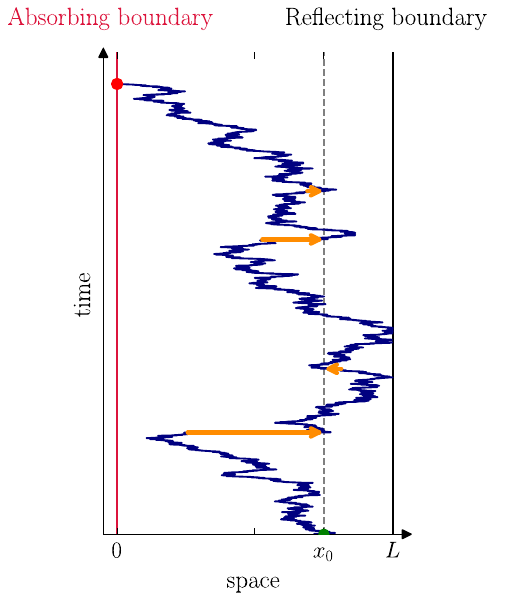}
    \caption{Schematic trajectory of an active Ornstein--Uhlenbeck particle on the interval $[0,L]$, with an absorbing boundary at $x=0$ and a reflecting boundary at $x=L$. Resetting returns the particle to $x_0$, and the propulsion velocity is reset according to the chosen protocol.}
    \label{fig:schematic}
\end{figure}

In this paper, we study the FPT statistics of an AOUP on a finite one-dimensional interval with an absorbing boundary at $x=0$ and a reflecting boundary at $x=L$, subject to Poissonian resetting (Fig.~\ref{fig:schematic}). We extend a perturbative framework, introduced by Wang \cite{Wang2021} to study the steady-state behaviour of weakly active particles near boundaries, to the first-passage problem. This yields analytical expressions for the Laplace-transformed survival probability, the FPT distribution, and its low-order moments. We first consider the problem without resetting and show that activity can either increase or decrease the MFPT, depending on the relative persistence and diffusive timescales. We then introduce resetting and determine both the onset of beneficial resetting and the finite range of resetting rates for which it lowers the MFPT of the active system. We also compare these results for different velocity resetting protocols. Finally, by comparing the active particle with resetting to the passive Brownian particle without resetting, we construct a phase diagram that identifies the regions of parameter space in which activity and resetting together reduce the MFPT.

%%%%%%%%%%%%%%%%%%%%%%%%%%%%%%
% Section: Governing equations and perturbative analysis %
%%%%%%%%%%%%%%%%%%%%%%%%%%%%%%
\section{Governing equations and perturbative analysis}
\label{sec:model}

%%%%%%%%%%%%%%%%%%%%%%%%%%%%%%
% Subsection: Model formulation and analytical framework %
%%%%%%%%%%%%%%%%%%%%%%%%%%%%%%
\subsection{Model formulation and analytical framework}
We consider an overdamped AOUP confined to the one-dimensional finite interval
$x(t)\in[0,L]$. In the absence of resetting, the position undergoes translational diffusion with coefficient $D$ and is advected by a self-propulsion velocity $v(t)$, modelled as an Ornstein--Uhlenbeck process with persistence time $\tau_p$ and noise strength $D_v$. The dynamics are governed by the Langevin equations
\begin{subequations}
\begin{align}
\frac{\dd x}{\dd t} &= v(t) + \sqrt{2D}\,\xi(t), \label{eq:dxdt}\\
\tau_p \frac{\dd v}{\dd t} &= -v(t) + \sqrt{2D_v}\,\eta(t), \label{eq:dvdt}
\end{align}
\end{subequations}
where $\xi(t)$ and $\eta(t)$ are independent Gaussian white noises with zero mean and delta-correlated variance.

Let $p(x,v,t)$ denote the joint probability density for the particle to have position $x$ and propulsion velocity $v$ at time $t$. The corresponding Fokker--Planck (FP) equation is
\begin{equation}
\frac{\partial p}{\partial t} = D\frac{\partial^2 p}{\partial x^2} - v\frac{\partial p}{\partial x} + \frac{1}{\tau_p}\frac{\partial}{\partial v}(vp) + \frac{D_v}{\tau_p^2}\frac{\partial^2 p}{\partial v^2}.
\label{eq:fp_dimensional}
\end{equation}
The boundary at $x=0$ is absorbing,
\begin{equation}
p(0,v,t)=0,
\end{equation}
and the boundary at $x=L$ is reflecting, corresponding to a no-flux condition in the $x$-direction,
\begin{equation}
J_x(L,v,t)=0,
\qquad
J_x= v\,p-D\,\frac{\partial p}{\partial x}.
\end{equation}

Following Ref. \cite{Wang2021}, we non-dimensionalise the Fokker--Planck equation via the rescalings
$x=\lambda\,\tilde x$, $v=\sigma\,\tilde v$ and $t=2\tau_p\,\tilde t$,
where $\lambda=\sqrt{2D\tau_p}$ is the diffusive length scale over the persistence time
and $\sigma=\sqrt{2D_v/\tau_p}$ is proportional to the steady-state root mean square propulsion velocity.
The dimensionless density
\begin{equation}
\tilde p(\tilde x,\tilde v,\tilde t) = \lambda\sigma\,p(\lambda\tilde x,\sigma\tilde v,2\tau_p\tilde t)
\end{equation}
then satisfies
\begin{equation}
\frac{\partial\tilde p}{\partial\tilde t} = \frac{\partial^2\tilde p}{\partial\tilde x^2} + \frac{\partial^2\tilde p}{\partial\tilde v^2} + 2\frac{\partial}{\partial\tilde v}(\tilde v\tilde p) - 2\epsilon\tilde v\frac{\partial\tilde p}{\partial\tilde x}.
\label{eq:fp_dimensionless}
\end{equation}
where
\begin{equation}
\epsilon=\sqrt{\frac{D_v}{D}}=\sqrt{\Pe},
\end{equation}
and $\Pe$ is the P\'eclet number which is the ratio of the rates of active and passive transport, and characterises the activity of the particle. Note that this is consistent with the usual definition ${\rm Pe} = v_0 \ell / D$, as the characteristic active velocity is here given by $v_0 \sim \sqrt{D_v/\tau_p}$ and the persistence length by $\ell \sim v_0 \tau_p = \sqrt{D_v \tau_p}$. For the remainder of this section, we omit the tildes for simplicity.

We are interested in obtaining the FPT statistics to absorption, given that the particle started at position $x_0$ and that the initial velocity density is $\rho_0(v)$. Let $S_0(t)$ denote the corresponding survival probability up to time $t$:
\begin{equation}
S_0(t)=\int_0^{L}\dd x\int_{-\infty}^{\infty}\dd v\, p(x,v,t),
\label{eq:S_integral}
\end{equation}
with $p(x,v,0)=\delta(x-x_0)\rho_0(v)$. The FPT density can be obtained directly from the survival probability
\begin{equation}
F(t)=-\frac{\partial S_0(t)}{\partial t}.
\label{eq:FPT_real}
\end{equation}
In Laplace space,
\begin{equation}
\hat F(s)=\int_0^\infty e^{-st}F(t)\,\dd t=1-s\hat{S}_0(s),
\end{equation}
and the moments of the dimensionless FPT distribution are given by
\begin{equation}
\langle T^n\rangle = (-1)^{n+1} \left. \frac{\partial^n}{\partial s^n} \big[s\hat{S}_0(s)\big] \right|_{s=0}.
\label{eq:moments_dimless}
\end{equation}
Hence, the FPT statistics can be obtained by solving the Fokker--Planck equation in Laplace space and computing the survival probability.

Taking the Laplace transform of the dimensionless FP equation \eqref{eq:fp_dimensionless} gives
\begin{equation}
s\hat p - p(x,v,0) = \frac{\partial^2\hat p}{\partial x^2} + \frac{\partial^2\hat p}{\partial v^2} + 2\frac{\partial}{\partial v}(v\hat p) - 2\epsilon\,v\frac{\partial\hat p}{\partial x}.
\label{eq:fp_dimensionless_laplace}
\end{equation}
where again we choose initial conditions
\begin{equation}
p(x,v,0)=\delta(x-x_0)\,\rho_0(v),
\qquad
\rho_0(v)=\frac{e^{-v^2}}{\sqrt{\pi}},
\label{eq:IC_dimless}
\end{equation}
i.e. a Gaussian initial velocity distribution.

%%%%%%%%%%%%%%%%%%%%%%%%%%%%%%%%%%%
% Subsection: Perturbative solution of the Fokker–Planck equation  %
%%%%%%%%%%%%%%%%%%%%%%%%%%%%%%%%%%%
\subsection{Perturbative solution of the Fokker–Planck equation}

Following Ref. \cite{Wang2021}, we consider a regular perturbation series expansion in the weak activity regime,
\begin{equation}
\hat p(x,v,s)=\sum_{n=0}^\infty \epsilon^n \hat p^{(n)}(x,v,s).
\end{equation}
Substituting into \eqref{eq:fp_dimensionless_laplace} yields
\begin{subequations}
\begin{align}
&s\hat p^{(0)} - \delta(x-x_0)\rho_0 = \mathcal L_0 \hat p^{(0)} + \frac{\partial^2 \hat p^{(0)}}{\partial x^2}, \label{eq:order0}\\
&s\hat p^{(n)} = \mathcal L_0 \hat p^{(n)} + \frac{\partial^2 \hat p^{(n)}}{\partial x^2} -2v\frac{\partial \hat p^{(n-1)}}{\partial x},\qquad n\ge1,
\label{eq:ordern}
\end{align}
\end{subequations}
where $\mathcal{L}_0$ is the Ornstein--Uhlenbeck operator acting in velocity space:
\begin{equation}
\mathcal L_0 f = \frac{\partial^2 f}{\partial v^2}+2\frac{\partial}{\partial v}(v f).
\end{equation}

Further simplification can be obtained by diagonalising $\mathcal L_0$ using Hermite polynomials. Specifically, we set
\begin{equation}
\hat p^{(n)}(x,v,s) = \sum_{m=0}^\infty A_m^{(n)}(x,s)\,\frac{e^{-v^2}}{\sqrt{\pi}}\,H_m(v),
\label{eq:Hermite_expansion}
\end{equation}
which reduces the problem to a set of coupled ODEs for the coefficients $A_m^{(n)}$.
At leading order,
\begin{equation}
\frac{\partial^2A^{(0)}_m}{\partial x^2}-(2m+s)A^{(0)}_m=-\delta(x-x_0)\delta_{m0},
\label{eq:A0_eq}
\end{equation}
and for $n\ge1$,
\begin{equation}
\frac{\partial^2A^{(n)}_m}{\partial x^2}-(2m+s)A^{(n)}_m = \frac{\partial}{\partial x}\Big[A^{(n-1)}_{m-1}+2(m+1)A^{(n-1)}_{m+1}\Big].
\label{eq:An_eq}
\end{equation}
The absorbing boundary at $x=0$ implies
\begin{equation}
A^{(n)}_m(0,s)=0\qquad m,n\ge0.
\end{equation}
At the reflecting boundary $x=L$, a no-flux condition must be imposed. At leading order, this implies
\begin{equation}
\left.\frac{\partial A_m^{(0)}}{\partial x}\right|_{x=L}=0,
\qquad m\geq 0,
\end{equation}
while for $n\ge1$,
\begin{equation}
A^{(n-1)}_{m-1}(L,s) +2(m+1)A^{(n-1)}_{m+1}(L,s) -\left.\frac{\partial A^{(n)}_{m}}{\partial x}\right|_{x=L} =0.
\label{eq:BCs}
\end{equation}
To leading order, only $A_0^{(0)}$ is non-zero. It then follows from Eq.\,\eqref{eq:An_eq} that $A_m^{(n)}$ can be non-zero only for $0\le m\le n$ with $n-m$ even. The first few non-zero coefficients are
\begin{equation}
A_0^{(0)};\quad
A_1^{(1)};\quad
A_0^{(2)},\,A_2^{(2)};\quad
A_1^{(3)},\,A_3^{(3)};\quad
A_0^{(4)},\,A_2^{(4)},\,A_4^{(4)}.
\end{equation}
In particular, the $m=0$ terms occur only at even orders.

At order $\mathcal{O}(1)$, the solution is given explicitly by the following piecewise function:
\begin{equation}
A_0^{(0)}(x,s)=
\begin{cases}
\frac{\sinh(\sqrt{s}\,x)\,\cosh(\sqrt{s}(L-x_0))}{\sqrt{s}\,\cosh(\sqrt{s}L)}, & 0\le x<x_0, \\[8pt]
\frac{\sinh(\sqrt{s}\,x_0)\,\cosh(\sqrt{s}(L-x))}{\sqrt{s}\,\cosh(\sqrt{s}L)}, & x_0<x\le L.
\end{cases}
\label{eq:A00_explicit}
\end{equation}
At order $\mathcal{O}(\epsilon)$, only $A_1^{(1)}$ does not vanish. The solution is again piecewise:
\begin{equation}
A_1^{(1)}(x,s)=
\begin{cases}
A_{1,-}^{(1)}(x,s), & 0\le x<x_0,\\[4pt]
A_{1,+}^{(1)}(x,s), & x_0<x\le L,
\end{cases}
\label{eq:A11_piecewise_rewrite}
\end{equation}
where
\begin{subequations}
\begin{align}
\begin{split}
A_{1,-}^{(1)}(x,s) &= \frac{1}{2}\Bigg[ \cosh\big((L-x_0)\sqrt{s}\big)\sech\big(L\sqrt{s}\big) \\
			   & \quad\quad\quad \times\Big(\cosh\big((L-x)\sqrt{s+2}\big) \sech\big(L\sqrt{s+2}\big) -\cosh\big(x\sqrt{s}\big) \Big)\\
			   &\quad\quad+\sinh\big(x\sqrt{s+2}\big)\sech\big(L\sqrt{s+2}\big) \\
			   &\quad\quad\quad\times \Big(\sinh\big((L-x_0)\sqrt{s+2}\big)+\sqrt{\frac{s+2}{s}}\,\sech\big(L\sqrt{s}\big)\sinh\big(x_0\sqrt{s}\big)\Big)\Bigg],
\end{split}\label{eq:A11_left} \\
\begin{split}
A_{1,+}^{(1)}(x,s) &= \frac{1}{2}\Bigg[\cosh\big((L-x)\sqrt{s+2}\big)\sech\big(L\sqrt{s+2}\big) \\
			   &\quad\quad\quad \times \Big(\cosh\big((L-x_0)\sqrt{s}\big)\sech\big(L\sqrt{s}\big)-\cosh\big(x_0\sqrt{s+2}\big)\Big) \\
			   &\quad\quad + \sinh\big(x_0 \sqrt{s}\big)\sech\big(L\sqrt{s}\big) \\ 
			   &\quad\quad\quad \times\Big(\sinh\big((L-x)\sqrt{s}\big) +\sqrt{\frac{s+2}{s}}\,\sech\big(L\sqrt{s+2}\big)\sinh\big(x\sqrt{s+2}\big)\Big) \Bigg].
\end{split}\label{eq:A11_right}
\end{align}
\end{subequations}
The same procedure can be continued recursively to obtain the higher order coefficients. The explicit expression for the second-order coefficient $A_0^{(2)}(x,s)$ is given in \ref{app:A20}.

The Laplace-transformed survival probability is
\begin{equation}
\hat S_0(s)=\int_0^L \dd x \int_{-\infty}^{\infty} \dd v\,\hat p(x,v,s).
\end{equation}
Substituting the Hermite expansion \eqref{eq:Hermite_expansion} and using
\begin{equation}
\int_{-\infty}^{\infty}\frac{e^{-v^2}}{\sqrt{\pi}}H_m(v)\,\dd v=\delta_{m,0},
\end{equation}
shows that only the $A^{(n)}_0$ coefficients contribute. Since the $m=0$ terms are non-zero only at even orders, the expansion of the survival probability contains only even powers of $\epsilon$:
\begin{equation}
\hat S_0(s) = \hat S_0^{(0)}(s) +\epsilon^2 \hat S_0^{(2)}(s) +\epsilon^4 \hat S_0^{(4)}(s) +\mathcal{O}(\epsilon^6).
\label{eq:S_expansion_clean}
\end{equation}
The leading-order Brownian term is
\begin{equation}
\hat S_0^{(0)}(s)
=
\frac{1-\cosh\big(\sqrt{s}(L-x_0)\big)\sech(L\sqrt{s})}{s}.
\label{eq:S0}
\end{equation}
The functions $\hat S_0^{(2)}$ and $\hat S_0^{(4)}$ then follow from the coefficients $A_0^{(2)}$ and $A_0^{(4)}$. The explicit form of $\hat S_0^{(2)}$ is given in \ref{app:S2}.

%%%%%%%%%%%%%%%%%%%%%%%%%%%%%%%
% Subsection: Renewal framework for stochastic resetting   %
%%%%%%%%%%%%%%%%%%%%%%%%%%%%%%%
\subsection{Renewal framework for stochastic resetting}

Having established the Laplace-transformed survival probability $\hat{S}_0(s)$ for the active process in the absence of resetting, we now embed these results into a renewal framework to include Poissonian resetting at rate $r$. At each reset, the particle position is returned to its initial value $x_0$ and a new propulsion velocity is drawn from the initial distribution $\rho_0(v)$. We denoted $S_0(t)$ the survival probability without resetting, let $S_r(t)$ denote the survival probability with resetting. The last renewal equation is given by
\begin{equation}
S_r(t) = e^{-rt}S_0(t) + r\int_0^t e^{-rt'}S_0(t')S_r(t-t')\,\dd t',
\label{eq:renewal_time}
\end{equation}
where the first term accounts for trajectories in which resetting does not occur, and the second term accounts for trajectories in which the last reset occurs at time $t-t'$. Taking the Laplace transform and solving for $\hat S_r(s)$ gives
\begin{equation}
\hat S_r(s) = \frac{\hat S_0(s+r)}{1-r\hat S_0(s+r)}.
\label{eq:Laplace_survival}
\end{equation}
Using the perturbative expansion $\hat S_0(s+r)=\sum_{n=0}^\infty \epsilon^n \hat S_0^{(n)}(s+r)$, we obtain the corresponding expansion for $\hat S_r$:
\begin{subequations}
\begin{align}
\hat S_r^{(0)} &= \frac{\hat S_0^{(0)}}{1-r\hat S_0^{(0)}},\\
\hat S_r^{(1)} &= \frac{\hat S_0^{(1)}}{1-r\hat S_0^{(0)}}
+ \frac{r\hat S_0^{(0)}\hat S_0^{(1)}}{\left[1-r\hat S_0^{(0)}\right]^2} = 0,\\
\hat S_r^{(2)} &=  \frac{\hat{S}_{0}^{(2)}}{1 - r\,\hat{S}_{0}^{(0)}}
   + \frac{r\left[\hat{S}_{0}^{(0)}\hat{S}_{0}^{(2)} + \left[\hat{S}_{0}^{(1)}\right]^{2}\right]}
          {\left[1 - r\,\hat{S}_{0}^{(0)}\right]^{2}}
   + \frac{r^{2}\hat{S}_{0}^{(0)}\left[\hat{S}_{0}^{(1)}\right]^{2}}
          {\left[1 - r\,\hat{S}_{0}^{(0)}\right]^{3}} = \frac{\hat{S}_{0}^{(2)}}
          {\left[1 - r\,\hat{S}_{0}^{(0)}\right]^{2}}.
\end{align}
\label{eq:S_r_expansion}
\end{subequations}
where for brevity we have dropped the explicit $(s+r)$ dependence in each of the expansion coefficients on the right-hand side of Eq. (\ref{eq:S_r_expansion}). We conclude that once $\hat S_0(s)$ is determined, the survival probability with resetting---and hence the corresponding FPT statistics---follows directly from Eq.\,\eqref{eq:Laplace_survival}.

%%%%%%%%%%%%%%%%%%%%%%%%%%%%%%
% Section: First-passage statistics and phase transitions   %
%%%%%%%%%%%%%%%%%%%%%%%%%%%%%%
\section{First-passage statistics and phase transitions}

We now use the perturbative framework developed in Sec.~\ref{sec:model} to determine how activity and stochastic resetting influence FPT statistics. We first consider the case with no resetting to isolate the effect of activity relative to passive Brownian motion. We then introduce resetting and determine both the onset of beneficial resetting and the range of resetting rates over which it lowers the MFPT of the active system. Finally, we compare the full active-resetting dynamics with the passive Brownian case without resetting, and determine the regions in parameter space for which the MFPT is reduced. For the remainder of this paper, to compare to the results of direct numerical simulations, we will redimensionalise our perturbative results. Throughout this section, we have set $L=\sqrt{2}$ and $D=1$, which effectively sets the diffusion timescale of the particle to unity. Consequently, all temporal quantities discussed below, most notably the persistence time $\tau_p$ and the typical resetting time $\tau_r \equiv 1/r$, should be interpreted as dimensionless ratios relative to this characteristic diffusion time.

\subsection{The isolated effect of activity on first-passage times}
\label{subsec:no_reset_results}

\begin{figure}
    \centering
    \includegraphics[width=\textwidth]{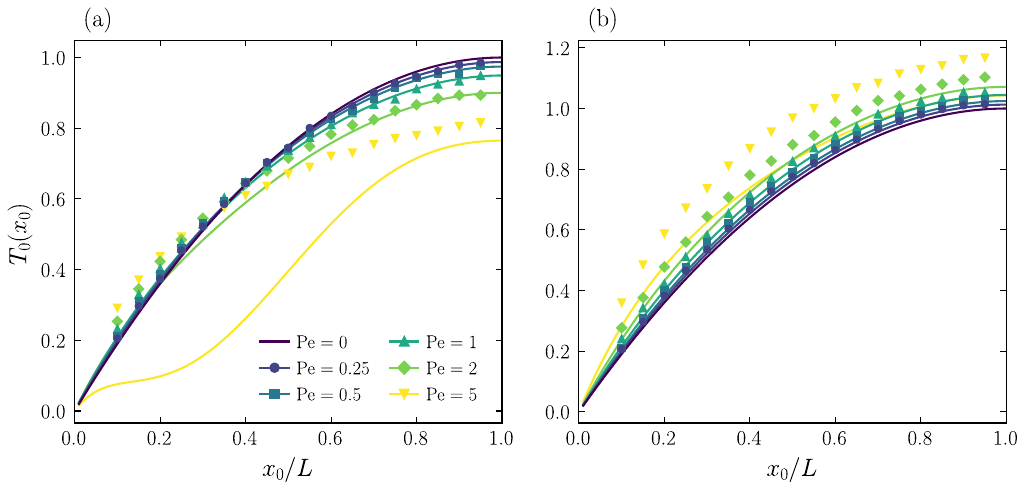}
    \caption{MFPT without resetting as a function of $x_0/L$ for (a) $\tau=0.25$ and (b) $\tau_p=1$ for the values of $\Pe$ indicated in the figure. Solid lines show the perturbative approximation to $\mathcal{O}(\Pe^2)$ and symbols show simulation results.}
    \label{fig:MFPT_no_resetting}
\end{figure}

First, we consider the dynamics in the absence of resetting ($r=0$). For symmetric initial velocity distributions
$\rho_0(v)=\rho_0(-v)$, all terms with odd orders in $\epsilon=\sqrt{\Pe}$ in our perturbative expansion are expected to vanish by symmetry. Thus, the MFPT admits the following expansion
\begin{equation}
T_0(x_0) = T_0^{(0)}(x_0) + \Pe\,T_0^{(2)}(x_0) + \Pe^2\,T_0^{(4)}(x_0) + \mathcal{O}(\Pe^3),
\label{eq:T0_expansion_results}
\end{equation}
The zeroth-order term corresponds naturally to the passive Brownian result given by
\begin{equation}
T_0^{(0)}(x_0)=x_0\left(L-\frac{x_0}{2}\right),
\label{eq:T00_result}
\end{equation}
and the first-order correction due to activity is given by
\begin{align}
T_0^{(2)}(x_0) = \sech\big(\sqrt{2}L\big)\Biggl( & \frac{1}{2} - Lx_0\nonumber +\left(\frac{x_0}{2}-L\right)x_0\cosh(\sqrt{2}L) - \frac{1}{2}\cosh\big(\sqrt{2}x_0\big) \nonumber \\
& + \frac{\sqrt{2}}{2}(L+x_0)\sinh(\sqrt{2}L) + \frac{\sqrt{2}}{2}L\sinh\bigl(\sqrt{2}(x_0-L)\bigr) \Biggr). 
\label{eq:T02_result}
\end{align}
The expression for $T_0^{(4)}$ is omitted here for simplicity, but is included where indicated in the results.

In Fig.\,\ref{fig:MFPT_no_resetting}, we show example plots of the MFPT, computed to order $\mathcal{O}(\Pe^2)$, as a function of $x_0/L$, which establishes the existence of two distinct regimes. For sufficiently large persistence times, activity increases the MFPT over the entire interval, indicating that persistent self-propulsion hinders absorption. For smaller persistence times, however, the effect depends on the starting position. Near the absorbing boundary, activity increases the MFPT, whereas further in the bulk it is lowered.

This behaviour is governed by the ratio of the persistence time to the diffusive time to absorption. A particle starting at $x_0$ reaches the absorbing boundary diffusively on a timescale $t_d(x_0)\sim x_0^2/D$. When $t_d(x_0)\ll \tau_p$, the particle maintains its initial propulsion velocity over the timescale for passive absorption. Trajectories initially directed away from the absorbing boundary can travel deep into the bulk, increasing the MFPT. When $t_d(x_0)\gg \tau_p$, the propulsion velocity decorrelates before diffusive absorption typically occurs, and the motion becomes effectively diffusive with an increased diffusivity. In this regime activity reduces the MFPT. The perturbative expansion displays these behaviours and agrees well with simulations up to $\Pe\approx 1$ (Fig.~\ref{fig:MFPT_no_resetting}). This indicates that the expansion remains a useful approximation for moderate activity.

\begin{figure}
    \centering
    \includegraphics[width=0.6\textwidth]{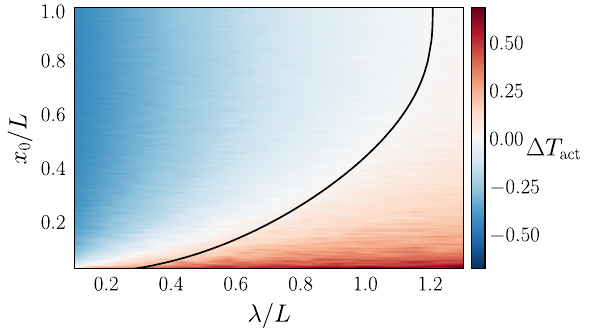}
    \caption{Crossover position $x_c/L$ as a function of the scaled persistence length $\lambda/L$. The colour map shows the activity-induced change in the MFPT, $\Delta T_{\mathrm{act}}$, obtained from simulations at $\Pe=1$, while the solid black curve shows the crossover $\Delta T_{\mathrm{act}}=0$ predicted by the perturbative theory.}
    \label{fig:xc_collapse}
\end{figure}

To characterise the crossover in behaviour at small persistence times, we consider the relative difference between the active and passive MFPTs,
\begin{equation}
\Delta T_{\mathrm{act}}(x_0;\Pe)
=
\frac{T_0(x_0;\Pe)-T_0^{(0)}(x_0)}{T_0^{(0)}(x_0)}.
\label{eq:deltaT}
\end{equation}
The crossover position $x_c$ is defined by $\Delta T_{\mathrm{act}}(x_c;\Pe)=0$.
Fig.~\ref{fig:xc_collapse} shows that the crossover position depends on $\tau_p$ through the ratio of the persistence length to the system size, and therefore takes the form
\begin{equation}
\frac{x_c}{L}
=
\mathcal{F}\!\left(\frac{\lambda}{L}\right),
\label{eq:xc_scaling_results_revised}
\end{equation}
where $\lambda=\sqrt{2D\tau_p}$ is the persistence length. As $\lambda/L$ increases, the crossover shifts to larger $x_0/L$, until beyond a threshold activity increases the MFPT for all initial positions.

\begin{figure}
    \centering
    \includegraphics[width=\textwidth]{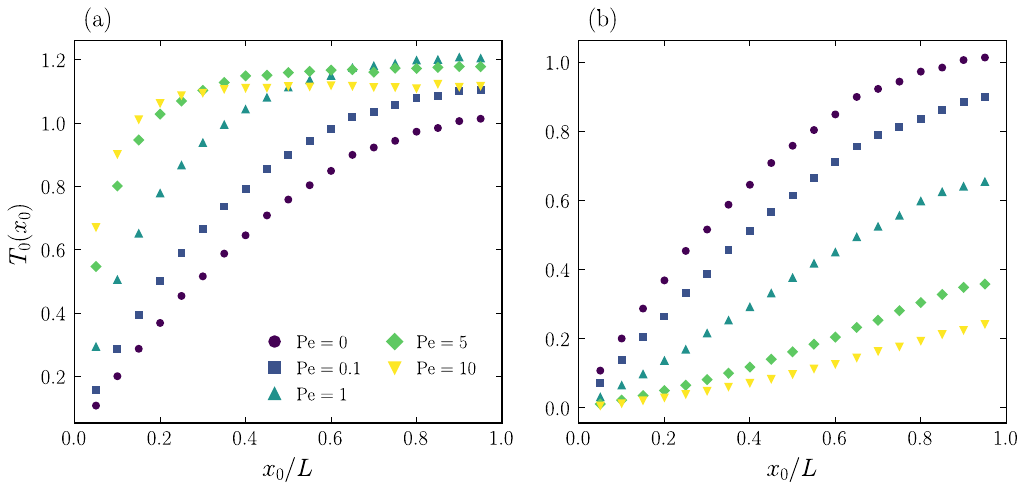}
    \caption{MFPT without resetting for fixed initial velocities, for (a) $v_0=+1$ and (b) $v_0=-1$ for $\tau_p=0.25$ and the values of $\Pe$ indicated in the figure. The results are obtained from simulations.}
    \label{fig:vp_initial}
\end{figure}

We next consider fixed initial velocities $v_0=\pm1$ instead of the symmetric Gaussian distribution. This isolates the effect of the initial propulsion direction on the MFPT. Since the perturbative theory developed in Sec. \ref{sec:model} was for the stationary Gaussian initial velocity distribution, the fixed-velocity results shown here are obtained from simulations only. Fig.~\ref{fig:vp_initial} shows that when $v_0=-1$, the particle is initially directed towards the absorbing boundary and activity lowers the MFPT throughout the interval. For $v_0=+1$, the initial propulsion is directed away from the target, and the crossover remains but is shifted to larger $x_0/L$.

\subsection{Onset of beneficial resetting and the monotonicity transition}
\label{sec:results_resetting}

\begin{figure}
    \centering
    \includegraphics[width=0.6\textwidth]{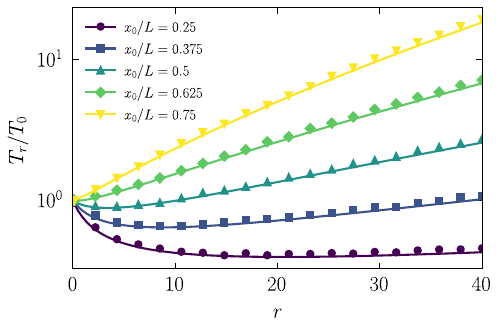}
    \caption{MFPT $T_r$ as a function of the resetting rate $r$ for varying values of $x_0/L$, at fixed $\tau_p=1$, and $\Pe=1$. The solid curves show the perturbative prediction to $\mathcal{O}(\Pe^2)$, and the symbols are the simulation results. Depending on the initial position, the dependence on $r$ is either monotonic or non-monotonic.}
    \label{fig:mfpt_vs_r}
\end{figure}

\begin{figure}
    \centering
    \includegraphics[width=\textwidth]{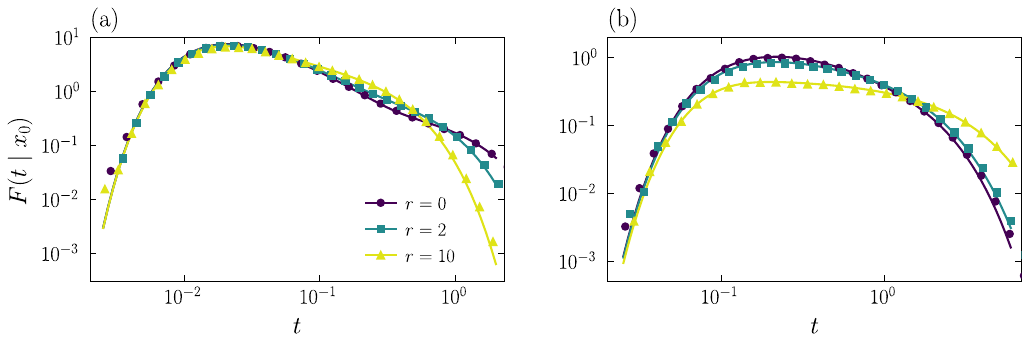}
    \caption{FPT distributions for (a) $x_0/L=0.25$ and (b) $x_0/L=0.75$ at fixed $\Pe=1$ and $\tau_p=1$. Solid lines and symbols denote the theory and simulations results respectively. Resetting suppresses the long-time tail when it is beneficial, but delays absorption when it interrupts trajectories that would otherwise reach the target.}
    \label{fig:FPT}
\end{figure}

We now introduce Poissonian resetting at rate $r$, with the particle returned to $x_0$ and the propulsion velocity redrawn from $\rho_0(v)$ at each reset. We then determine how the MFPT $T_r$, calculated to $\mathcal{O}(\Pe^2)$, varies with $r$ at fixed persistence time $\tau_p$ and P\'eclet number $\Pe$. In Fig.~\ref{fig:mfpt_vs_r}, we show that the response depends on the initial position, as previously found for a passive Brownian particle confined to a finite interval \cite{Pal2017}. For particles starting sufficiently close to the absorbing boundary, $T_r$ is non-monotonic in $r$, and an optimal resetting rate exists. For larger $x_0/L$, the dependence becomes monotonic and resetting only increases the MFPT. Resetting is therefore only beneficial when it suppresses long excursions into the bulk without frequently interrupting trajectories that would otherwise reach the absorbing boundary. The perturbative prediction agrees well with simulations across the range of resetting rates shown. In \ref{app:relative_error}, we report the relative error between theory and simulations for $r=10$. This distinction is also visible in the first-passage-time distributions shown in Fig.~\ref{fig:FPT}, where resetting either suppresses or enhances the long-time tail depending on the chosen initial position.

The boundary between these two regimes is determined by the behaviour of $T_r$ near $r=0$. Resetting is initially beneficial if
\[
\left.\frac{dT_r}{dr}\right|_{r=0}<0,
\]
or equivalently if the coefficient of variation of the no-resetting first-passage process exceeds unity \cite{Pal2017}. In terms of the Laplace-transformed survival probability, this criterion becomes
\begin{equation}
-\hat S_0'(0)>\hat S_0(0)^2,
\label{eq:beneficial_hatS_resetting_results}
\end{equation}
where the prime denotes differentiation with respect to the Laplace variable
$s.$
Substituting the perturbative expansion of $\hat S_0$ into this condition gives the phase boundary separating monotonic from non-monotonic resetting response.

\begin{figure}
    \centering
    \includegraphics[width=0.6\textwidth]{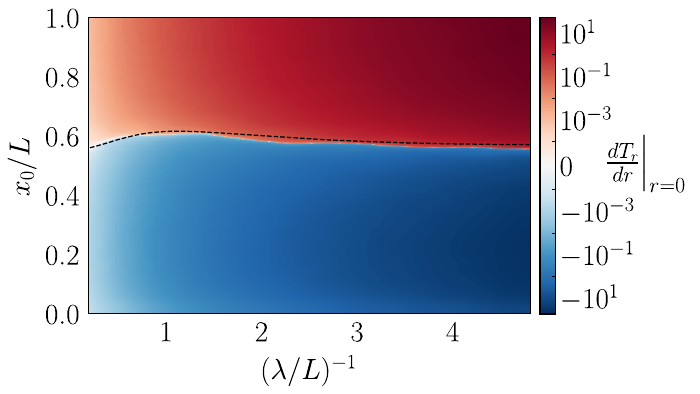}
    \caption{Phase diagram separating monotonic and non-monotonic dependence of the MFPT on the resetting rate, for $\Pe=0.6$. The colour map shows $\left.\frac{dT_r}{dr}\right|_{r=0}$ obtained from simulations by measuring the coefficient of variation for the non-resetting process, and the dashed curve is the analytic prediction for the transition.}
    \label{fig:phase_diagram}
\end{figure}

The phase diagram for this transition is shown in Fig.~\ref{fig:phase_diagram}. In the passive Brownian limit, the critical position is \cite{Pal2017}
\begin{equation}
\frac{x_0^*}{L}=1-\frac{\sqrt{5}}{5}\approx0.55.
\end{equation}
Activity shifts this boundary to larger $x_0/L$ values, so that resetting remains beneficial for particles starting further from the absorbing boundary than in the passive case. We interpret this as follows: each reset reinitialises both the particle position and the propulsion velocity leading to some of the trajectories being reset in the direction of the absorbing boundary.
This criterion identifies where resetting is initially beneficial, but not the range of resetting rates for which it lowers the MFPT.

\subsection{Range of beneficial resetting rates}
\label{subsec:resetting_window}

\begin{figure}
    \centering
    \includegraphics[width=0.6\textwidth]{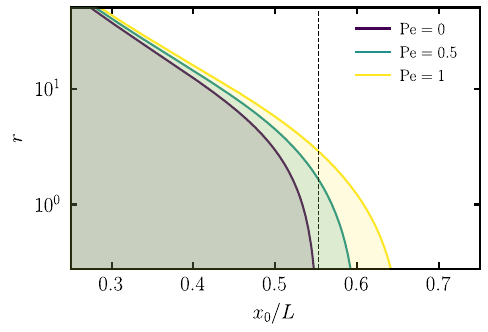}
    \caption{Region in the $(x_0/L,r)$ plane for which resetting lowers the MFPT relative to the no-resetting active case, for varying $\Pe$ at fixed $\tau_p=1$. For each $\Pe$, the shaded region shows the parameter range where resetting is beneficial. The vertical dashed line shows the value of the critical position for the passive Brownian case. The beneficial-resetting region grows as the activity is increased.}
    \label{fig:beneficial_resetting_region}
\end{figure}

To determine this range, we compare $T_r$ with the no-resetting active MFPT $T_0$. For fixed $\tau_p$, resetting lowers the MFPT whenever
\begin{equation}
T_r(x_0,r;\Pe)<T_0(x_0;\Pe).
\label{eq:beneficial_region_condition}
\end{equation}
This inequality defines a region in the $(x_0/L,r)$ plane within which resetting is beneficial. Fig.~\ref{fig:beneficial_resetting_region} shows that resetting is beneficial only up to a finite resetting rate $r_c(x_0)$. Beyond this point, resetting interrupts trajectories too often, and the MFPT exceeds the no-resetting value. As $x_0\to x_0^{*-}$, the range of beneficial resetting rates shrinks continuously and $r_c\to0$, consistent with the monotonicity transition in $T_r$. Additionally, $x_0^*$ increases with $\Pe$, so that the region of the $(x_0/L,r)$ plane in which resetting is beneficial becomes larger as the activity is increased.

\subsection{Influence of the velocity reinitialisation protocol}
\label{subsec:reset_protocol}
We now consider how the phase boundary for the monotonicity transition depends on the propulsion velocity resetting protocol. In this section, the initial and reset velocities are fixed to $v_0=\pm1$. As the perturbative framework developed in Sec.~\ref{sec:model} relies on a stationary Gaussian initial velocity distribution, the analytical approach is not immediately applicable to fixed-velocity protocols. Therefore, the results presented in this section are obtained via direct numerical simulations. 

\begin{figure}
    \centering
    \includegraphics[width=\textwidth]{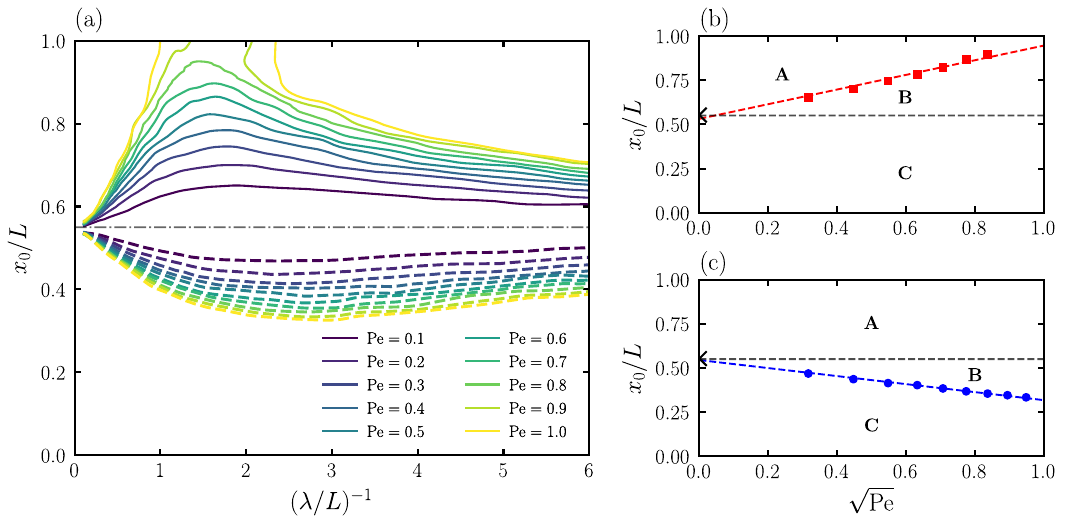}
    \caption{Effect of the velocity-resetting protocol on the monotonicity transition. (a) Phase boundaries in the $((\lambda/L)^{-1},x_0/L)$ plane for different values of $\Pe$. Solid curves correspond to $v_0=-1$, while dashed curves correspond to $v_0=+1$. The horizontal dash-dotted line shows the passive Brownian value. (b,c) Extremal values of the initial position $x_0/L$ as a function of $\sqrt{\Pe}$ for $v_0=-1$ in (b) and $v_0=+1$ in (c). Symbols show simulation results, with the passive Brownian limit marked by a black cross. The dashed lines are linear regressions including the Brownian limit and show the approximately linear dependence on $\sqrt{\Pe}$. Regions A, B, and C indicate, respectively, parameter regimes in which the MFPT is non-monotonic in $r$ for all $(\lambda/L)^{-1}$, non-monotonic only over a restricted range of $(\lambda/L)^{-1}$, and monotonic in $r$ for all $(\lambda/L)^{-1}$.}
    \label{fig:reset_protocol_combined}
\end{figure}

In Fig.\,\ref{fig:reset_protocol_combined}(a),  we show that the phase boundary is shifted relative to the passive Brownian case. Resetting with $v_0=-1$ shifts the boundary to larger values of $x_0/L$, whereas resetting with $v_0=+1$ shifts it to smaller values of $x_0/L$. The corresponding extremal values of the critical initial position are shown in Figs.~\ref{fig:reset_protocol_combined}(b) and~\ref{fig:reset_protocol_combined}(c). These extrema vary approximately linearly with $\sqrt{\Pe}$, consistent with the typical distance travelled before the propulsion velocity decorrelates which is of order $\sigma \tau_p$.

For sufficiently large $\Pe$, resetting with $v_0=-1$ can be beneficial for all initial positions. The regions labelled $\mathbf{A}$, $\mathbf{B}$, and $\mathbf{C}$ in Figs.~\ref{fig:reset_protocol_combined}(b)-(c) distinguish the different monotonicity regimes in the $(x_0/L,\Pe)$ plane. At fixed P\'eclet number, in region $\mathbf{A}$ the MFPT is non-monotonic in $r$ for all values of $(\lambda/L)^{-1}$, whereas in region $\mathbf{C}$ the dependence is monotonic for all values of $(\lambda/L)^{-1}$. In region $\mathbf{B}$, there exists an optimal resetting rate over a restricted range of $(\lambda/L)^{-1}$.

\subsection{MFPT minimisation and the optimal resetting rate}

\begin{figure}
    \centering
    \includegraphics[width=0.6\textwidth]{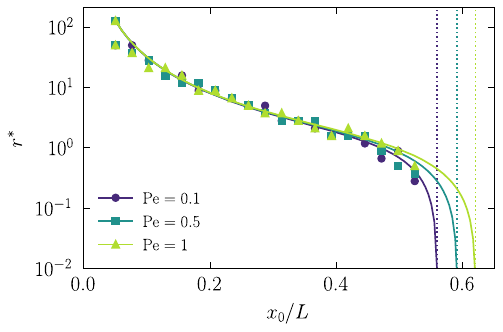}
    \caption{Optimal resetting rate $r^*$ as a function of $x_0/L$ for $\tau_p=1$. Solid lines and symbols denote the theory and simulations results respectively for varying $\Pe$.}
    \label{fig:rstar_vs_x0}
\end{figure}

Within the beneficial-resetting regime, the MFPT exhibits a minimum at a finite resetting rate $r^*$. Fig.~\ref{fig:rstar_vs_x0} shows $r^*$ as a function of $x_0/L$ for varying P\'eclet number. An optimum only exists when $x_0<x_0^*$, and as $x_0\to x_0^{*-}$, $r^*$ approaches zero continuously, consistent with the transition to monotonic behaviour at the phase boundary.

\subsection{Active versus passive search: combined phase diagrams}

The previous sections considered how activity changes the FPT statistics in the absence of resetting, and how resetting modifies the active process. We now compare the active particle with resetting to the passive Brownian particle without resetting.

We define
\begin{equation}
T_{\mathrm{rel}}(x_0,r,\tau_p)
=
\frac{T_r(x_0,r,\tau_p)}{T_0^{(0)}(x_0)},
\end{equation}
where $T_r$ is the MFPT of the active particle with resetting and $T_0^{(0)}$ is the passive Brownian MFPT without resetting. The condition $T_{\mathrm{rel}}=1$ defines the boundary between the region in which activity and resetting lower the MFPT and the region in which they increase it.

\begin{figure}
    \centering
    \includegraphics[width=\textwidth]{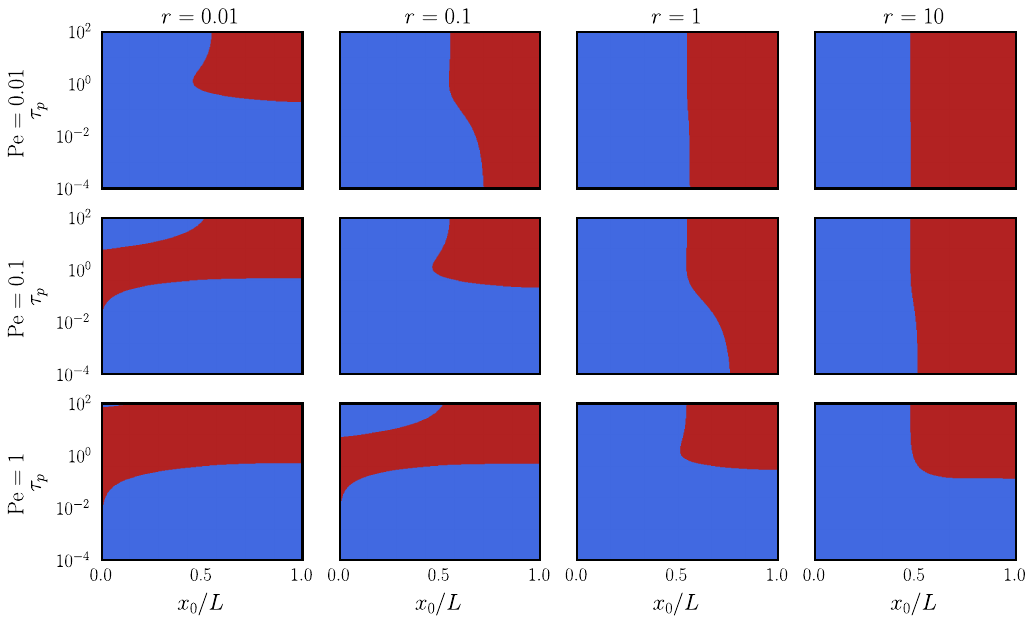}
    \caption{Phase diagram in the $(x_0/L,\tau_p)$ plane for different values of the resetting rate $r$ and the P\'eclet number $\Pe$. The blue region corresponds to $T_{\mathrm{rel}}<1$, where the active Ornstein--Uhlenbeck particle with resetting has a lower MFPT than the passive Brownian particle without resetting, and the red region corresponds to $T_{\mathrm{rel}}>1$. }
    \label{fig:full_phase_diagram}
\end{figure}

Fig.~\ref{fig:full_phase_diagram} shows the phase diagram in the $(x_0/L,\tau_p)$ plane for different values of $r$ and $\Pe$. In the absence of resetting, the boundary between the two regions reduces to the crossover found in Sec.~\ref{subsec:no_reset_results}. Once resetting is introduced, whether resetting is beneficial or not depends on the competition between three timescales: the diffusive absorption time $t_d(x_0)\sim x_0^2/D$, the persistence time $\tau_p$, and the mean reset time $\tau_r=1/r$.

For small $x_0/L$, the effect of resetting is controlled by the distance the particle typically travels before either its propulsion velocity decorrelates or the next reset occurs. This active distance is set by the minimum of the persistence and reset timescales,
\begin{equation}
\ell_{\mathrm{act}}\sim \sigma \min(\tau_p,\tau_r).
\end{equation}
At fixed $\Pe$, this gives $\ell_{\mathrm{act}}\sim \sqrt{\tau_p\,\Pe}$ when $\tau_p\ll \tau_r$, whereas for $\tau_p \gg \tau_r$ the reset interval becomes the limiting timescale and $\ell_{\mathrm{act}}\sim \sqrt{\Pe}/(r\sqrt{\tau_p})$.

For small $\Pe$, this distance remains too small for persistence to be strongly detrimental and the active-resetting process has the lower MFPT. As $\Pe$ increases, the dependence on $\tau_p$ becomes non-monotonic. When $\tau_p$ is small, the active displacement remains limited and the active-resetting process remains favoured. At intermediate $\tau_p$, the active displacement becomes large enough for trajectories directed away from the boundary to travel sufficiently deep into the bulk, and the passive Brownian process becomes more efficient. For larger $\tau_p$, the reset interval becomes the shorter timescale, $\ell_{\mathrm{act}}$ decreases and the active-resetting process is again favoured.

For larger $x_0/L$, resetting repeatedly returns the particle to a position farther from the absorbing boundary, interrupting trajectories before they can make sufficient progress towards the target. In this case, whether activity lowers the MFPT depends on how the persistence time compares with the diffusive absorption time. When $\tau_p \ll t_d$, the motion is effectively diffusive, and for sufficiently large $\Pe$ the active-resetting process can still have a lower MFPT. As $\tau_p$ increases, persistent motion increasingly delays arrival at the absorbing boundary, and the passive Brownian process becomes more efficient.

As $r$ is increased further, the region for which $T_{\mathrm{rel}}<1$ becomes smaller overall, since frequent restarting eventually prevents the particle from making sufficient progress between reset events. In the large-$r$ limit, the Brownian particle has the lower MFPT throughout the parameter space.

\section{Conclusion \& outlook}
\label{sec:conclusion}

In this paper, we have studied the FPT statistics of an active Ornstein--Uhlenbeck particle on a finite interval with an absorbing boundary at one end, a reflecting boundary at the other, subject to stochastic resetting. Working in the weak-activity regime, we solved the Fokker--Planck equation perturbatively and then used renewal theory to obtain analytical expressions for the Laplace-transformed survival probability, the FPT density, and the corresponding low-order moments. Comparison with simulations shows that our perturbation theory remains accurate for  surprisingly moderate activity up to P\'eclet numbers of order one.

In the absence of resetting, activity was found to either increase or decrease the MFPT depending on the relative magnitudes of the persistence time and the diffusive absorption time. In particular, for sufficiently short persistence times, there is a crossover in the initial position that separates regions in which activity delays absorption from those in which it accelerates it.

When stochastic resetting is introduced, its effect on the MFPT is again parameter-dependent. Depending on the initial position, resetting can either lower the MFPT and lead to an optimal resetting rate, or increase it monotonically. We identified the boundary between these two regimes, and showed that when resetting is initially beneficial it remains so only up to a given resetting rate.

Finally, by comparing the active particle with resetting to the passive Brownian particle without resetting, we constructed phase diagrams identifying the regions in which activity and resetting together lower the MFPT. Overall, the FPT behaviour is controlled by the competition between diffusive, persistence, and resetting timescales.

There are several avenues for future work. Since the present analysis is restricted to the weak activity regime, an important next step is to determine how the crossover in the no-resetting problem and the phase boundaries for beneficial resetting are modified in the strongly persistent regime, where large activity effects may become important. A second direction is to study alternative reset protocols analytically. Our numerical results already show that the resetting transition strongly depends on how the propulsion velocity is reset, and it would therefore be useful to further study alternative reset protocols such as fixed-velocity resetting or one in which the position is reset while the propulsion velocity is left unchanged. Furthermore, in biologically motivated search problems, target encounter does not necessarily lead to immediate absorption. For this reason, extension to include partially absorbing targets would be important for bringing the model closer to realistic search processes.

\appendix
\section{Expression for the second-order coefficient $A_0^{(2)}$}
\label{app:A20}

The second-order coefficient $A_0^{(2)}$ is given by
\begin{equation}
A_0^{(2)}(x,s)=
\begin{cases}
A_{0,-}^{(2)}(x,s), & 0\leq x\leq x_0,\\[3pt]
A_{0,+}^{(2)}(x,s), & x_0\leq x\leq L.
\end{cases}
\label{eq:A20_piecewise}
\end{equation}
where the two branches read
\begin{subequations}
\begin{align}
A_{0,-}^{(2)}(x,s) ={}& \alpha_-(s,x_0) \left[\sqrt{s}e^{-\sqrt{s}\,x} - \sqrt{s}\cosh\!\bigl(\sqrt{s+2}\,x\bigr)\right] \nonumber \\
                      &+\beta_-(s,x_0)\frac{e^{-\sqrt{s}(x_0+L)}}{e^{L\sqrt{s}} + e^{-L\sqrt{s}}}\sech\!\bigl(L\sqrt{s+2}\bigr)\sinh(\sqrt{s}\,x) \nonumber \\
                      &+\frac{1}{2}\gamma_-(s,x) \sech(L\sqrt{s})\cosh\!\bigl(\sqrt{s}(L-x_0)\bigr), \label{eq:A20minus}\\[6pt]
A_{0,+}^{(2)}(x,s) ={}& \alpha_+(s,x_0)\left[\sqrt{s}e^{\sqrt{s}(L-x)} -\sqrt{s+2}\, \sinh\!\bigl(\sqrt{s+2}(L-x)\bigr) \right] \nonumber\\
                      &+\beta_+(s,x_0)\frac{e^{-\sqrt{s}x_0}}{e^{L\sqrt{s}} + e^{-L\sqrt{s}}}\sech\!\bigl(L\sqrt{s+2}\bigr)\cosh\!\bigl(\sqrt{s}(L-x)\bigr) \nonumber \\
                      &+\frac{1}{2}\gamma_+(s,x)\sech(L\sqrt{s})\sinh(\sqrt{s}\,x_0)
\label{eq:A20plus}
\end{align}
\end{subequations}

In doing so, we have defined the following coefficient functions
\begin{subequations}
\begin{align}
\alpha_-(s,x_0)={}&-\frac{1}{2}\sqrt{\frac{s+2}{s}}\sech\!\bigl(L\sqrt{s+2}\bigr)\Bigg[\sinh\!\bigl(\sqrt{s+2}\,(L-x_0)\bigr) \label{eq:alpha_minus_def}\\
&+\sech(L\sqrt{s})\left(e^{-L\sqrt{s+2}}\cosh\!\bigl(\sqrt{s}(L-x_0)\bigr)+\sqrt{\frac{s+2}{s}}\,\sinh(\sqrt{s}\,x_0)\right)\Bigg],\nonumber\\[6pt]
\alpha_+(s,x_0)={}&-\frac{1}{2}\sech\!\bigl(L\sqrt{s+2}\bigr)\Bigg[\cosh\!\bigl(\sqrt{s+2}\,x_0\bigr)\label{eq:alpha_plus_def}\\
&-\sech(L\sqrt{s})\left(\cosh\!\bigl(\sqrt{s}(L-x_0)\bigr)+e^{L\sqrt{s+2}}\sqrt{\frac{s+2}{s}}\,\sinh(\sqrt{s}\,x_0)\right)\Bigg].\nonumber\\
\gamma_-(s,x)={}&-\frac{s+2}{\sqrt{s+2}}\,e^{-\sqrt{s+2}\,x} -\left(\frac{L}{2}-\sqrt{s+2}\right)e^{-\sqrt{s}\,x} +\left(\frac{L}{2} -x\right)\cosh(\sqrt{s}\,x)\label{eq:gamma_minus_def}\\[6pt]
\gamma_+(s,x)={}&\frac{s+2}{\sqrt{s}}\,e^{\sqrt{s+2}\,(L-x)} -\left(\frac{L}{2}+\sqrt{s+2}\right)e^{\sqrt{s}\,(L-x)} -\left(\frac{L}{2} -x\right)\sinh(\sqrt{s}\,(L-x)) \label{eq:gamma_plus_def} 
\end{align}
\end{subequations}

and 
\begin{subequations}
\begin{align}
\beta_-(s,x_0)={}&e^{(L + x_0)\sqrt{s}}\sqrt{s}\cosh\!\bigl(\sqrt{s+2}\,x_0\bigr)-e^{\sqrt{s}x_0}\sqrt{s+2}\sinh\!\bigl(\sqrt{s+2}(L-x_0)\bigr)\nonumber\\
			  &-e^{\sqrt{s}x_0}\frac{2\sinh(\sqrt{s}\,x_0)}{\sqrt{s}}\Bigg[(1+s)\sech(L\sqrt{s})+\sqrt{s(s+2)}\sinh\!\bigl(L\sqrt{s+2}\bigr)\tanh(L\sqrt{s})\nonumber\\
			  &-\frac{1}{4}\cosh\!\bigl(L\sqrt{s+2}\bigr)\Big[L\sqrt{s}\,\big(1+2\tanh(L\sqrt{s})\big)-2\big(1+s+\sqrt{s}\,x_0\big)\Big]\Bigg] \nonumber\\
			  &+\sqrt{s+2}\sinh\!\bigl(L\sqrt{s+2}\bigr)-e^{L\sqrt{s}}\sqrt{s} + x_0 \cosh\!\bigl(L\sqrt{s+2}\bigr)\sinh(L\sqrt{s})e^{L\sqrt{s}} \nonumber\\
			  &-e^{L\sqrt{s}} \cosh\!\bigl(L\sqrt{s+2}\bigr)\cosh(L\sqrt{s}) \left(\frac{1+s}{\sqrt{s}}+\frac L2\right)
\label{eq:beta_minus_def}\\[6pt]
\beta_+(s,x_0)={}&e^{(L + x_0)\sqrt{s}}\sqrt{s}\cosh\!\bigl(\sqrt{s+2}\,x_0\bigr)-e^{\sqrt{s}x_0}\sqrt{s+2}\sinh\!\bigl(\sqrt{s+2}(L-x_0)\bigr)\nonumber\\
			  &-e^{\sqrt{s}x_0}\frac{2\sinh(\sqrt{s}\,x_0)}{\sqrt{s}}\Bigg[(1+s)\sech(L\sqrt{s})+\sqrt{s(s+2)}\sinh\!\bigl(L\sqrt{s+2}\bigr)\tanh(L\sqrt{s})\nonumber\\
			  &-\frac{1}{4}\cosh\!\bigl(L\sqrt{s+2}\bigr)\Big[L\sqrt{s}\,\big(1+2\tanh(L\sqrt{s})\big)-2\big(1+s+\sqrt{s}\,x_0\big)\Big]\Bigg] \nonumber\\
			  &+\sqrt{s+2}\sinh\!\bigl(L\sqrt{s+2}\bigr)-e^{L\sqrt{s}}\sqrt{s}-x_0\cosh\!\bigl(L\sqrt{s+2}\bigr) \nonumber\\
\label{eq:beta_plus_def}
\end{align}
\end{subequations}

\section{Expression for $\hat S_0^{(2)}$}
\label{app:S2}

The second-order correction to the survival probability in the absence of
resetting is

\begin{equation}
\hat S_0^{(2)}=
\frac{\sech^2(L\sqrt{s})\sech(L\sqrt{s+2})}{2s}A(s,x_0),
\label{eq:S^2_0_def}
\end{equation}

where
\begin{align}
A(s,x_0) ={} &s\cosh(L\sqrt{s})\left[\cosh(\sqrt{s}\,x_0)-\cosh(\sqrt{s+2}\,x_0)\right] \nonumber\\
		   &-\sqrt{s}\,x_0\cosh(L\sqrt{s})\cosh(L\sqrt{s+2})\sinh\!\bigl(\sqrt{s}(L-x_0)\bigr)\nonumber\\
		   &+\sqrt{s(s+2)}\,\sinh(L\sqrt{s})\Big[\cosh\!\bigl(\sqrt{s}(L-x_0)\bigr)\sinh(L\sqrt{s+2})\nonumber\\
		   &\hspace{4.2cm}-\cosh(L\sqrt{s})\sinh\!\bigl(\sqrt{s+2}(L-x_0)\bigr)\Big]\nonumber\\
		   &+\sinh(\sqrt{s}\,x_0)\Big[-L\sqrt{s}\cosh(L\sqrt{s+2})\nonumber \\
		   &\hspace{2cm}-2(1+s)\sinh(L\sqrt{s})+\sqrt{s(s+2)}\,\sinh(L\sqrt{s+2})\Big].
\label{eq:A_def}
\end{align}

\section{Relative error of the perturbative approximation}
\label{app:relative_error}

\begin{figure}
    \centering
    \includegraphics[width=0.6\textwidth]{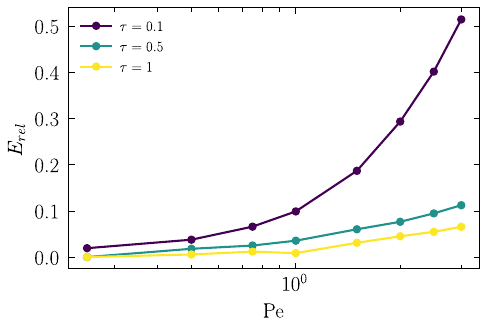}
    \caption{Relative error between the perturbative approximation to $\mathcal{O}(\Pe^2)$ and simulation results as a function of $\Pe$, for $r=10$ and $x_0/L=0.5$.}
    \label{fig:relative_error_appendix}
\end{figure}

To quantify the accuracy of the perturbative approximation, we compute the relative error between the analytic and simulated MFPTs,
\begin{equation}
E_{\mathrm{rel}}
=
\frac{\left|T_{\mathrm{theory}}-T_{\mathrm{sim}}\right|}{T_{\mathrm{sim}}}.
\end{equation}
In Fig.\,\ref{fig:relative_error_appendix}, we show $E_{\mathrm{rel}}$ as a function of $\Pe$ at $r=10$ and fixed initial position $x_0/L=0.5$, for different values of the persistence time $\tau_p$. For the range shown, the error decreases as $\tau_p$ increases. At fixed $r$, increasing $\tau_p$ reduces the size of the active displacement between resets. The effect of activity is therefore suppressed, and the dynamics remain closer to the Brownian limit.

\end{document}